\documentclass{article}

\usepackage[preprint]{format}

\usepackage[utf8]{inputenc} %
\usepackage[T1]{fontenc}    %
\usepackage{hyperref}       %
\usepackage{url}            %
\usepackage{booktabs}       %
\usepackage{amsfonts}       %
\usepackage{xcolor}         %
\usepackage{nicefrac}       %
\usepackage{microtype}      %
\usepackage{xcolor}         %

\usepackage{bm}
\usepackage{hyperref}
\usepackage{url}
\usepackage{multirow}
\usepackage{graphicx}
\usepackage{algorithmicx}
\usepackage{algorithm}
\usepackage{algpseudocode}
\usepackage{wrapfig}
\usepackage{multirow}
\usepackage{amsmath}
\usepackage{amssymb}
\usepackage{booktabs}
\usepackage{array} 
\usepackage{adjustbox}
\usepackage{bbding}
\usepackage{ulem}
\usepackage{makecell}
\usepackage{enumitem}
\usepackage{colortbl}
\usepackage{wrapfig}
\PassOptionsToPackage{numbers, compress, sort}{natbib}
\definecolor{codegreen}{rgb}{0,0.3,0.6}
\definecolor{codegray}{rgb}{0.5,0.5,0.5}
\definecolor{codepurple}{rgb}{0.58,0,0.82}
\definecolor{backcolour}{rgb}{0.95,0.95,0.92}

\definecolor{darkblue}{rgb}{0.0,0.0,0.66}   
\hypersetup{colorlinks=true,linkcolor=red,citecolor=darkblue}

\title{Antigen-specific Antibody Multi-modal Foundation Model for Functional Antibody Design}

\author{%
  Xiaoliang Shi\footnotemark[1],\ \ 
  Zichen Wang\thanks{Equal contribution.},\ \ 
  Runze Ma,\ \ 
  Zhongyue Zhang,\ \ 
  Shuangjia Zheng\thanks{Corresponding author.} \\[4pt]
  Shanghai Jiao Tong University \\
  \texttt{shuangjia.zheng@sjtu.edu.cn}
}

\begin{document}

\maketitle

\begin{abstract}
Antibodies are essential proteins that play a central role in immune recognition by binding specific antigen molecules. Although recent protein language models have enabled progress in single-chain protein modeling and generation, they often fall short in antigen-specific antibody design, where effective modeling requires explicit pairing between antibody and antigen, particularly at the epitope level.
To address these limitations, we introduce AAMFM, an \textbf{A}ntigen-specific \textbf{A}ntibody \textbf{M}ultimodal \textbf{F}oundation \textbf{M}odel that learns unified representations of antibody sequences and structures conditioned on antigen context. AAMFM incorporates rich antigen information including geometric interfaces and epitope annotations via a cross-modal adapter, enabling joint modeling of antibody-antigen interactions in a shared latent space. To further guide the model toward functional relevance, we fine-tune AAMFM using Calibrated Direct Preference Optimization (Cal-DPO), leveraging preference signals extracted from a strong structural prior (AlphaFold3) to align learning with binding-specific objectives.
Extensive experiments demonstrate that AAMFM achieves state-of-the-art performance in functional antibody design, revealing its potential for antigen-specific antibody engineering. Our code is available at 
\url{https://github.com/XL-S224/AAMFM}.
\end{abstract}

\section{Introduction}

As fundamental proteins within the immune system, antibodies play a crucial role in identifying and neutralizing specific molecules known as antigens. The precise targeting ability of antibodies predominantly originates from their Complementarity Determining Regions (CDRs), which are the key determinants of binding affinity to these antigens \citep{jones1986replacing,ewert2004stability,xu2000diversity,akbar2021compact}. Given that therapeutic antibodies represent a major and rapidly expanding class of large molecule drugs, the ability to effectively design functional and plausible CDRs and structure fit with the corresponding antibody is paramount for creating novel and potent treatments. 

However, developing such antibodies often depends on labor-intensive experimental approaches like large antibody library screening\citep{breitling1991surface,hoogenboom2005selecting}, which frequently falls short in effectively designing antibodies that target to specific antigen. While computational approaches, particularly deep learning-based Protein Language Models (PLMs) \citep{lin2023evolutionary,nijkamp2023progen2,ruffolo2024design,hayes2025simulating}, have shown promise in accelerating protein engineering, applying them to antibody design faces challenges. Many existing PLMs primarily model sequence information and struggle in protein complexes, hindering end-to-end optimization problem and limiting their ability to fully capture the complex interplay between sequence, structure, and residue interaction. Thus, general PLMs might struggle to generate highly specific antibodies for a given antigen, particularly at the epitope level. Conversely, specialized antibody sequence-structure co-design models \citep{jiniterative2021,luo2022antigen,kongMEAN,kong2023end,zhuabx,wang2024iggm} are often constrained by limited available data, which can lead to model overfitting and thus hampering their generalization. Moreover, their lack of general protein knowledge may result in the generation of implausible protein sequence or structure.

Beyond the challenge of generation, a critical limitation in current research lies in the evaluation and guidance of antibody design\citep{zhou2024antigen,ye2024proteinbench,renmulti}. Traditional energy-based scoring functions such as Rosetta \citep{alford2017rosetta} exhibit low correlation with experimental binding affinities \citep{luo2023rotamer}, making them relatively unreliable. Recent breakthroughs in protein structure foundation models, such as AlphaFold3 (AF3) \citep{abramson2024accurate}, have significantly expanded the modeling capabilities to multi-chain protein complexes. Notably, AF3 provides structure confidence scores like the ipTM and ranking score (AF3 score), which have been shown to correlate with binding interface quality \citep{lu2024alphafold3,hitawala2024has,wee2024benchmarking}. These provide a promising signal for assessing and refining antibody-antigen interactions. We identify an additional challenge: how can we effectively leverage such strong structural prior to steer antibody generation towards more functional, antigen-specific solutions?

To address these challenges, we introduce AAMFM, an Antigen-specific Antibody Multimodal Foundation Model, designed for unified antibody sequence-structure representation learning and antigen-specific design. Built upon the powerful foundation multi-modal protein language model ESM3 \citep{hayes2025simulating}, AAMFM is first adapted to the antibody domain using large-scale sequence-structure antibody pairs and subsequently fine-tuned on high-resolution experimental antibody-antigen complex data. Our model distinctively integrates antibody sequence, structure, and crucial antigen-related information—including its geometric features and specific epitope sites—into a shared latent space using a cross-modal adapter. A key innovation lies in enhancing antigen-specific functional antibody generation through preference optimization. We evaluate candidate antibodies generated by the model by predicting their complex structure with the target antigen using a reproduced version (Protenix \citep{bytedance2025protenix}) of AF3. The resulting quality metrics, specifically the AF3 score which reflects both the sequence's intrinsic foldability and its predicted binding capability to the antigen, are then used to define preferences. AAMFM is further fine-tuned on the preference data using Cal-DPO, aligning the model towards generating antibodies that satisfy these critical structural and binding criteria.

Our contributions are as follows: (1) We propose \textbf{AAMFM}, an Antigen-specific Antibody Multimodal Foundation Model that unifies antibody sequence, structure, and antigen information to enable antibody representation and design. 
(2) We construct an \textbf{antibody preference dataset} consisting of about 30k antibody sequences with corresponding AF3 predicted structures and annotated preference labels. The dataset is built by labeling each sequence using AF3's score and the pseudo log-likelihood (PLL) score from the antibody language model.
(3) We introduce a novel \textbf{AF3-based preference optimization} framework to guide antigen-specific functional and plausible antibody generation. By leveraging AF3-based scoring reflecting both foldability and binding affinity, AAMFM is fine-tuned using Cal-DPO to generate high-quality, target-specific antibodies.
(4) Experimental results show \textbf{state-of-the-art} performance on multiple antibody design tasks, highlighting the effectiveness of combining multimodal foundation model, antigen-specific adapter and preference optimization for therapeutic antibody development.

\section{Related Work}

\textbf{Protein Multimodal Language Models} 
Protein Language Models (PLMs) have made significant strides in recent years by learning from large-scale protein sequence data to effectively capture protein interactions, structures, and functions \citep{meier2021language, rao2021msa,elnaggar2021prottrans,ferruz2022protgpt2,lin2023evolutionary,nijkamp2023progen2,ruffolo2024design}. Recently, the research trend has shifted from unimodal to multimodal approaches, enhancing model capabilities by incorporating additional structural or functional information. LMDesign \citep{zheng2023structure} integrates geometric information into language models using an adapter, demonstrating superior performance compared to sequence-only models. ESM3 \citep{hayes2025simulating}, a representative large-scale multimodal model, is jointly pre-trained on vast amounts of protein sequence, structure, and function data, enabling the generation of protein data across different modalities. DPLM-2 \citep{wangdplm} extends discrete diffusion protein language model to the multimodal domain, focusing on learning the joint distribution of sequences and structures. In specific applications, S$^2$ALM\citep{yin2024s} is an antibody-specific language model combining sequence and structure; however, it can't design antibody binding to specific antigen. In contrast to these methods, our approach utilizes the powerful base model and fine-tunes it using multi-level finetune strategy. This strategy aims not only to achieve accurate antibody representation but also to effectively perform antigen-specific functional antibody design.

\textbf{Computational Antibody Design}
Early antibody design methods were often limited to energy-based approaches \citep{lapidoth2015abdesign,adolf2018rosettaantibodydesign}. Recently, deep learning approaches mainly follow two directions: antibody sequence design and sequence-structure co-design. 
The methods used for antibody sequence design include language models \citep{ruffolo2021deciphering,olsen2022ablang,wang2023pre,gao2023pre} and inverse folding models \citep{dreyer2023inverse,hoie2024antifold,wang2024retrieval}. 
Antibody sequence-structure co-design methods mainly taking antibody-antigen complex as a graph and using graph networks, or using diffusion-based model co-design structure and sequence of antibody CDR \citep{jiniterative2021,kongMEAN,kong2023end,lin2024geoab, luo2022antigen,zhuabx,martinkus2024abdiffuser,zhou2024antigen,wang2024iggm,renmulti}. 
Albeit powerful, existing co-design models are often constrained by the limited availability of training data, which makes it challenging for them to fully learn and capture the deep, complex features in protein. On the other hand, while language models generally possess strong representational capabilities, they typically exhibit weaker performance in generating functional antibodies specifically targeted specifically to a given antigen. To overcome these limitations, we propose a novel multimodal antibody language model.

\section{Methods}
\subsection{Preliminary and task formulation}
\begin{figure}[t]
    \centering
    \includegraphics[width=1\linewidth]{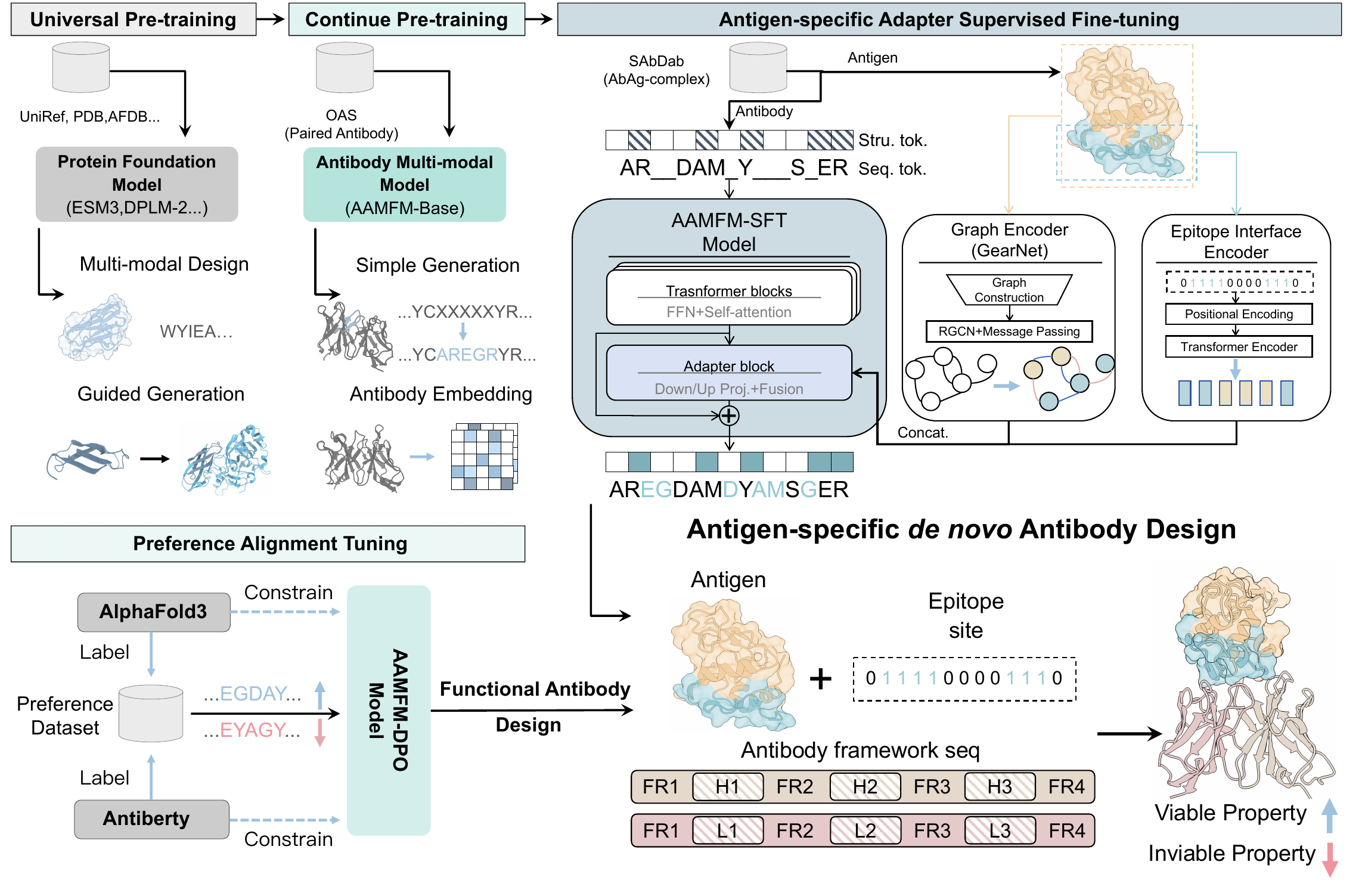}
    \vspace{-3mm} 
    \caption{Hierarchical training and inference pipeline of the proposed model, including pre-training, adapter-based SFT and preference optimization. During the inference phase, the model takes the antigen, epitope, and antibody framework region sequence as input, and outputs the antibody CDRs sequence and the complete antibody structure.
} %
    \label{fig1}
    \vspace{-4mm}
\end{figure}

\textbf{ESM3} ESM3 is a large-scale, multimodal generative language model designed to jointly model and generate protein sequence, structure, and function \citep{hayes2025simulating}. A key methodological innovation of ESM3 is the discrete tokenization of all input modalities.
While sequence is tokenized per amino acid, 3D atomic structure is uniquely represented by encoding local atomic neighborhoods around each residue into discrete tokens using a Vector Quantized Variational Autoencoder (VQ-VAE) \citep{van2017neural}. Let \( z \) denote the structure tokens, obtained as:

\begin{equation}
    z = \mathrm{quantize} \left( \mathrm{encode}_{\text{struct}}\left(\mathrm{knn}_{k=16}(x_{C_\alpha}),\ T \right) \right)
\end{equation}

where  $x_{C_\alpha} \in \mathbb{R}^{L \times 3}$  denotes the backbone $C_\alpha$ coordinates and  $T \in SE(3)^L$  the corresponding rigid transformations. The encoder processes the spatial context of each residue by selecting its 16 nearest neighbors and aggregates their relative geometric features before quantization. The encoded structure token can be inversely decoded to residue coordinates.

These distinct token tracks are embedded and processed within a unified latent space by the transformer blocks. ESM3 is trained using a generative masked language modeling (MLM) objective across all modalities. Unlike standard MLM with a fixed masking rate, ESM3 employs a noise schedule, sampling different masking rates during training. This allows the model to learn the conditional probabilities required for generating any modality from any combination of others, enabling flexible iterative generation from arbitrary prompts. %

\textbf{Task formulation} 
An antibody is composed of a heavy chain and a light chain. The terminal portion of each heavy-light chain pair forms a complementary site that binds to a specific antigen epitope. The precise recognition and interaction surface within this site is formed by six key loops known as Complementarity-Determining Regions (CDRs): three on the heavy chain (CDR-H1, H2, H3) and three on the light chain (CDR-L1, L2, L3) \citep{presta1992antibody, al1997standard}. The remaining regions typically form conserved framework regions. Therefore, our focus is on designing the complete CDR sequences and the full antibody structure. Given the antigen information $F_{ag}$, epitope position $I$, and the antibody framework sequence $s_{f}$, the model aims to jointly design the sequences of all six CDRs $s_{c}$ and the full antibody structure $X$.

\subsection{Multi-level pretraining and finetune strategies}

As shown in Figure \ref{fig1}, we structure the model's learning process in multiple stages. Initially, the model is trained on general protein sequences and structures, learning broad protein representations and basic generation patterns (for details, please refer to ESM3 \citep{hayes2025simulating}). However, antibody sequences and structures exhibit significant discrepancies compared to general proteins, such as their unique immunoglobulin fold domain and the highly variable CDRs that dictate their specificity. To enable the model to better understand and generate functional antibodies, we design the subsequent hierarchical post-training and fine-tuning strategies:

\textbf{Antibody Domain Adaptation Pre-training} Addressing the gap between the general model and the antibody domain, we first perform initial domain adaptation using a large-scale dataset specific to antibodies. Specifically, we additionally train the model using approximately 1.4 million paired antibody sequence-structure data from the Observed Antibody Space (OAS) database \citep{olsen2022ablang}. The structure of the corresponding sequences are predicted by AbodyBuilder2 \citep{abanades2023immunebuilder} and IgFold \citep{ruffolo2023fast}, which has been proved useful in previous works \citep{dreyer2023inverse,hoie2024antifold}. The core objective of this stage is to familiarize the pre-trained model with antibody-specific sequence and general structure patterns, and the fundamental sequence-structure correspondence, thereby adapting it to the antibody biological space and laying the foundation for more refined downstream tasks.

\textbf{Antigen-specific Supervised Fine-tuning and Preference Optimization} 
After the model has been initially adapted to the antibody domain, we introduce higher-quality data for supervised fine-tuning (SFT). This stage utilizes data from Structural Antibody Database (SAbDab) \citep{dunbar2014sabdab}, which contains a large number of high-resolution antibody structures determined by experimental methods, along with their corresponding authentic sequences. The experimental structural data helps the model learn more precise atomic-level interactions and conformational details. Considering that the ultimate goal of antibody design is often binding to a specific antigen, we incorporate antigen information through an adapter during this phase (section \ref{adapter}).  To further enhance the quality of generated antibodies and align the model better with functional and plausible design objectives, we introduce an additional fine-tuning stage using a DPO-style method (section \ref{DPO}).

\textbf{Training Objective}
We formulate the training objective similar to ESM3. Unlike general-purpose protein language models, antibodies exhibit a unique structural pattern: highly conserved framework regions and highly diverse CDRs. Applying a uniform masking probability across all positions fails to capture this distinction and may underemphasize the CDRs, which are functionally critical. To better capture these characteristics, we introduce a dynamic masking strategy tailored to different training stages.
During post-training, we apply random masking across all positions with 70 $\%$ probability, and with the remaining 30$\%$, we mask all CDR residues while keeping the framework regions fully visible. During supervised fine-tuning, we adopt a 50\% probability for masking across all positions, a 25\% probability for fully masking the CDR regions, and a 25\% probability for applying masking only within the CDRs.

The mask $m$ is applied to the input tokens $x$ (representing sequence or structure tokens), and the model learns to predict the original identity of the masked tokens $x_i$ given the unmasked context $x_{\setminus m}$. The training objective $\mathcal{L}$ is defined as:
\begin{equation}
    \mathcal{L} = -\mathbb{E}_{x,m} \left[ \frac{1}{|m|} \sum_{i \in m} \log p(x_i | x_{\setminus m}) \right]
\end{equation}

\subsection{Antigen Geometric-Epitope-aware Adapter}
\label{adapter}
To integrate antigen geometric and antigen-antibody interface information in designing the antigen-specific antibody, we propose the Antigen Geometric-Epitope-aware Adapter. This module is designed as a lightweight adapter \citep{houlsby2019parameter} integrated via a residual connection \citep{he2016deep} after the output $h$ of the final attention block of the model, which 
contains only 5.1M parameters ($\sim$0.36\% of the 1.4B-parameter base model). It computes an adjustment term $\Delta h$ to produce a refined output $h' = h + \Delta h$. The designed adapter primarily consists of the following modules:

\begin{wrapfigure}[33]{r}{0.5\textwidth}
        \centering
        \vspace{-4mm}
        \includegraphics[width=0.5\textwidth]{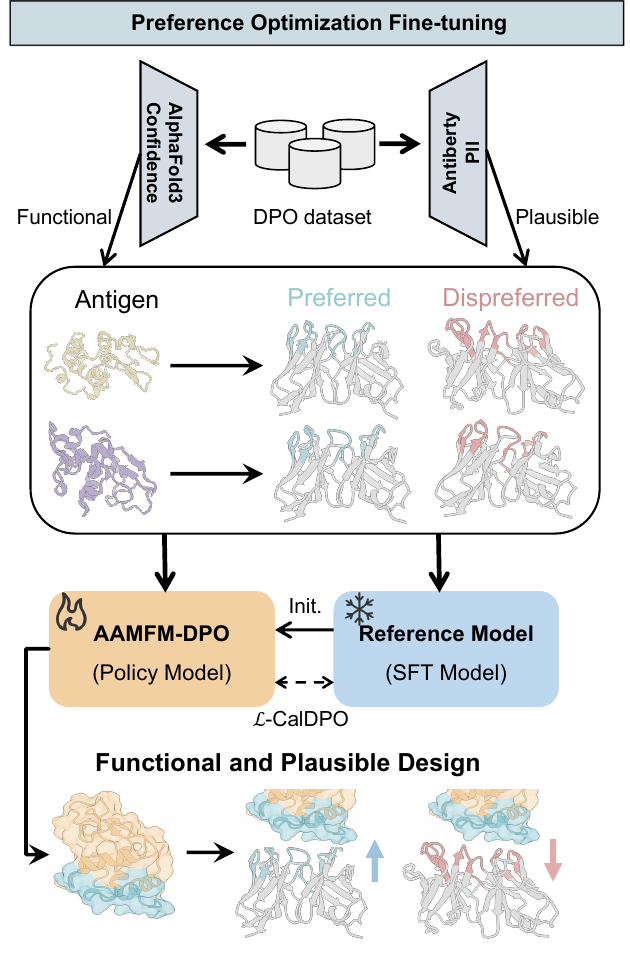}
        \vspace{-3mm}
        \caption{Illustration of the dataset curation process and the Cal-DPO framework. The contrastive-like approach aligns the model toward generating preferred functional sequences with higher probability. }
        \label{dpo_fig}
\end{wrapfigure}

\textbf{Antigen Feature Encoder} The antigen graph $G_{ag}$ is input into GearNet \citep{zhangprotein} and the output antigen geometric node feature $f_{{geo}}$ is then processed using this module. 
To address the variability in antigen lengths, a convolutional network projects the input features to a bottleneck dimension, yielding a geometric antigen representation $ h_{\text{ag}} = \mathrm{AntigenEncoder}(f_{{geo}})$.
If the complex does not have antigen ($f_{{geo}}$ is not provided), a learnable default feature is used instead.

\textbf{Epitope Interface Encoder} This module processes binary information $I$ (with 1 indicating epitope regions and 0 for non-epitope regions) representing the epitope site on the antigen. Similar to previous work \citep{deng2024nanobody}, the interface information is first embedded and combined with learnable positional encodings. Subsequently, a Transformer encoder layer captures contextual information within the interface region, outputting position-wise interface features $h_{\text{ep}} = \mathrm{InterfaceEncoder}(I)$.

\textbf{Feature Fusion} The backbone features $h$ are down-projected, then concatenated with the processed antigen geometry and interface features  $h_{\text{ag}}$, $h_{\text{ep}}$. The concatenated result is fused via an MLP. The output feature $M$ is mapped back to the original dimension via an up-projection layer, yielding the final adjustment term $\Delta h$.
This adjustment is added to the original input via a residual connection.

This module effectively incorporates the antigen's geometric information and epitope interface positions into the model's representations, enhancing the antigen specificity of the model in antibody design tasks, thereby enabling the generation of antibodies with better functionality.

\subsection{AlphaFold3-based Preference Optimization}
\label{DPO}

\textbf{Training Objective} Reinforcement Learning from Human Feedback (RLHF) \citep{ouyang2022training} methods have demonstrated strong capabilities in aligning language models with human preferences across various domains. Algorithms such as Direct Preference Optimization (DPO) \citep{rafailov2023direct,widatalla2024aligning} simplifies RLHF by using contrastive learning on preference data, and has proven effective. 

As shown in Figure \ref{dpo_fig},
to align our model with desired properties reflected in preference data, specifically focusing on CDRs, we employ Cal-DPO \citep{xiao2024cal}. While DPO effectively optimizes for relative preferences between sequences, it focuses mainly on the relative values of implicit rewards associated with two responses while ignoring their actual values, resulting in suboptimal alignment with human preferences.
Cal-DPO addresses this limitation by incorporating an additional calibration loss, encouraging the model’s output confidence to better reflect actual sequence quality (a comparison of preference optimization methods is provided in Appendix B).

Let $\pi_{\theta}$ denote the policy model (the model being fine-tuned) and $\pi_{ref}$ be a frozen reference model (the frozen SFT model). 
As in standard DPO, the training dataset consists of preference tuples, training data consists of tuples $(c, y_w, y_l)$, where $c$ represents the input context (masked antibody sequence $s_{fr}$, antigen features $G_{ag}$, and interface information $I$), $y_w$ is the preferred CDR sequence, and $y_l$ is the rejected CDR sequence.

The standard DPO loss is formulated as:

\begin{equation}
\mathcal{L}_{\text{DPO}}(\theta; \pi_{\text{ref}}) = - \mathbb{E}_{(c, y_w, y_l) \sim \mathcal{D}} \left[ \log \sigma \left( \beta \left( \hat{r}_\theta(c, y_w) - \hat{r}_\theta(c, y_l) \right) \right) \right]
\end{equation}

where $\sigma$ is the sigmoid function, $\beta$ is the DPO temperature parameter, and
\begin{equation}
\hat{r}_\theta(c, y) = \log \frac{\pi_\theta(y|c)}{\pi_{\text{ref}}(y|c)}
\end{equation}
\vspace{-3mm}

represents the log-likelihood ratio between policy model and reference model. 

In our implementation, the log-likelihoods $\log \pi_\theta(y|c)$ and $\log \pi_{\text{ref}}(y|c)$ are both computed only over the tokens within the CDR regions instead of the whole sequence. Besides, we normalize the log-likelihoods by the number of CDR tokens to mitigate bias towards longer CDRs. 

Cal-DPO introduces an additional calibration loss term. Defining the target margin $M = 1 / (2\beta)$, the calibration loss encourages the log-likelihood ratios to align with this margin:
\begin{equation}
\mathcal{L}_{\text{Cal}}(\theta; \pi_{\text{ref}}) = \mathbb{E}_{(c, y_w, y_l) \sim \mathcal{D}} \left[ \left( \hat{r}_\theta(c, y_w) - M \right)^2 + \left( \hat{r}_\theta(c, y_l) + M \right)^2 \right]
\end{equation}
\vspace{-3mm}

The final training loss is a weighted sum of the DPO and calibration terms (in our implementation, with a weighting factor $\lambda=1$ for $\mathcal{L}_{\text{Cal}}$):
\vspace{-3mm}

\begin{equation}
\mathcal{L}_{\text{Cal-DPO}} = \mathcal{L}_{\text{DPO}} + \lambda \cdot \mathcal{L}_{\text{Cal}}
\end{equation}

\textbf{Preference Definition} 
To construct the preference dataset, we first sample candidate sequences from the SFT model and deduplicate. Each sequence is then scored by Protenix~\citep{bytedance2025protenix}, a reproduction of AlphaFold3, to obtain a structure-based estimate of antigen-binding confidence at the designed CDR--antigen interface, which we use as the primary oracle for antibody functionality. Relying solely on this score, however, risks reward hacking: the model may learn to exploit the structural oracle, producing sequences with inflated predicted binding scores that deviate from the natural antibody distribution. To counteract this, we introduce pseudo-log-likelihood (PLL) under the antibody language model AntiBERTy~\citep{ruffolo2021deciphering} as a sequence-level plausibility constraint, measuring compatibility with natural antibody statistics. Crucially, PLL is only weakly correlated with the AF3 score on our preference dataset (Spearman $\rho = 0.13$), confirming that it acts as a nearly orthogonal signal rather than a redundant one.

We therefore define preferences via a dual-threshold rule: given a candidate pair $(s^+, s^-)$, $s^+$ is labeled as preferred only if it simultaneously exceeds $s^-$ by at least $0.2$ in AF3 score and $0.1$ in PLL; pairs failing either condition are discarded. This strict dual-threshold criterion provides a clear, unambiguous preference signal while preventing the model from exploiting either metric in isolation.

\section{Experiments}

\begin{table}[t]%
    \centering
    \caption{Results of full antibody design on SAbDab dataset}
    \begin{tabular}{cccccc}
    \toprule
         Method&  Pll $\uparrow$ &AF3-score $\uparrow$ & PHR $\downarrow$  &pTM $\uparrow$ &ipTM $\uparrow$\\
    \midrule
      ESM3-open & -1.40&0.846 & \textbf{0.400 }&0.876  & 0.839\\
         Diffab& -1.36& 0.850 &  0.445 &0.881 &0.842\\
 dyMEAN& -1.23&  0.842&  0.465 &0.874 & 0.840\\
 AbX& -0.96&  0.862  &0.480  &\uline{0.892} &0.855\\

\midrule
\textbf{ AAMFM-SFT}& \uline{-0.91}&\uline{0.870} &   0.465 &0.890 & \uline{0.865}\\
 
\textbf{AAMFM-CalDPO} & \textbf{-0.87}&\textbf{0.892 }&  \uline{0.443} & \textbf{0.908}&\textbf{0.888} \\

   \bottomrule
 \end{tabular}

    \label{tab2}
\end{table}

To evaluate the generative capabilities of our model, we conduct experiments on two primary tasks: \textit{de novo} antibody co-design and epitope-binding antibody CDR-H3 co-design. In addition, we perform comprehensive ablation studies to assess the contributions of each component. Experiment details and ablation are provided in Section~\ref{sec:ablation}.

\subsection{\textit{De novo} Antibody Sequence-Structure Co-design}
\label{codesign}

\begin{table}[t]
\vspace{-2mm}
\caption{Results of CDRs design on SAbDab dataset}
\begin{center}  
\label{tab1}
\arrayrulecolor{black}
\resizebox{.96\textwidth}{!}{
\begin{adjustbox}{width=\textwidth}%
\fontsize{15}{15}\selectfont
\begin{tabular}{cccccccccc}
\toprule
\multirow{2}{*}{Method} & \multicolumn{2}{c}{CDR-H1} & \multicolumn{2}{c}{CDR-H2} & \multicolumn{2}{c}{CDR-H3} \\
                        & AAR(\%) $\uparrow$ & RMSD $\downarrow$ & AAR(\%) $\uparrow$ & RMSD $\downarrow$ & AAR(\%) $\uparrow$ & RMSD $\downarrow$ \\
\midrule
ESM3-open  & 67.06 & 0.94 & 49.54 & \uline{0.89} & 32.19 & 3.31 \\
Diffab     & 62.32 & 0.92 & 44.82 & 0.96         & 32.33 & 3.15 \\
dyMEAN     & 75.71 & 1.09 & 66.67 & 0.98         & 38.46 & 3.51 \\
AbX        & 79.92 & \textbf{0.85} & 69.85 & \textbf{0.76} & \textbf{43.24} & 2.80 \\
\midrule
\textbf{AAMFM-SFT}    & \textbf{81.94} & 0.90          & \uline{70.86}  & 0.90 & \uline{40.32} & \textbf{2.54} \\
\textbf{AAMFM-CalDPO} & \uline{80.12}  & \uline{0.87}  & \textbf{71.59} & 0.92 & 39.88         & \uline{2.65}  \\
\midrule
\midrule
\multirow{2}{*}{Method} & \multicolumn{2}{c}{CDR-L1} & \multicolumn{2}{c}{CDR-L2} & \multicolumn{2}{c}{CDR-L3} \\
                        & AAR(\%) $\uparrow$ & RMSD $\downarrow$ & AAR(\%) $\uparrow$ & RMSD $\downarrow$ & AAR(\%) $\uparrow$ & RMSD $\downarrow$ \\
\midrule
ESM3-open  & 49.83 & 0.82          & 51.72 & 0.70         & 44.43 & 1.27 \\
Diffab     & 48.07 & 0.86          & 52.22 & 0.56         & 43.31 & 1.39 \\
dyMEAN     & 69.86 & \uline{0.71}  & 80.00 & \textbf{0.41}& 57.78 & 1.79 \\
AbX        & \textbf{79.37} & \textbf{0.70} & \uline{82.96} & \uline{0.42} & 65.96 & 1.18 \\
\midrule
\textbf{AAMFM-SFT}    & 75.57          & 0.95 & \textbf{83.87} & 0.73 & \textbf{72.70} & \uline{1.11}  \\
\textbf{AAMFM-CalDPO} & \uline{75.89}  & 1.02 & 82.09          & 0.74 & \uline{72.26}  & \textbf{1.10} \\
\bottomrule
\end{tabular}
\end{adjustbox}
}
\end{center}
\end{table}

\vspace{-3mm}

\textbf{Setup} For model training, we adopt the open-source version of ESM3-open (1.4B) as the base model. 
The first-stage training is performed on a subset of OAS dataset for 2 epochs. 
In the second-stage supervised fine-tuning, we utilize a curated version of the SAbDab dataset containing experimentally determined antibody-antigen complex structures, training the model for 45 epochs to learn precise antigen-specific interactions.
Following prior work \citep{luo2022antigen, zhuabx}, we filter out structures with resolution lower than 4Å and exclude antibodies that target non-protein antigens. We cluster antibodies based on 50\% CDR-H3 sequence identity and strictly filter our training data against the RAbD dataset  \citep{adolf2018rosettaantibodydesign}  to prevent data leakage.
The resulting fine-tuned model is referred to as \textbf{AAMFM-SFT}. We then perform preference alignment fine-tuning using Cal-DPO. We first sample 13,000 candidate CDR sequences using the AAMFM-SFT model. These sequences are evaluated by Protenix and AntiBERTy.
For each antigen, candidate sequences are combinatorially paired and filtered by the dual-threshold criterion, yielding approximately 30k preference pairs in total, which are used for training over 4 epochs.
The resulting model is referred to as \textbf{AAMFM-CalDPO}. 

For model inference, the model takes antigen and the antibody framework sequence as input, and generates the sequences of all six CDRs along with the full antibody structure. Notably, only dyMEAN and our method require only the antibody framework sequence as input, while other methods rely on antibody framework structure.

\textbf{Baselines} For antibody sequence-structure co-design, we compare AAMFM with a series of methods, including: \textbf{ESM3} \citep{hayes2025simulating}, a multi-modal protein foundation model; \textbf{Diffab} \citep{luo2022antigen},\textbf{AbX} \citep{zhuabx}, generative diffusion models; \textbf{dyMEAN} \citep{kong2023end}, a graph discriminative model.

\textbf{Metrics} We evaluate our designs by assessing both individual CDRs and the complete antibody. 
For the complete antibody sequence and structure, our evaluation includes: (1) the pseudo-log-likelihood (Pll) calculated by AntiBERTy to gauge sequence plausibility; (2) the AF3 score, derived from predicting the complex of the designed antibody and antigen, which comprehensively reflects the design's foldability and its antigen-binding functionality; (3) the AF3 pTM, representing the average predicted confidence for each chain's structure when the designed antibody and antigen are co-folded, indicating design foldability and plausibility; (4) the AF3 ipTM, specifically measures the predicted binding functionality between the designed antibody and the antigen; (5) the proportion of hydrophobic residues (PHR), which assess the antibodies specificity designed by the models.

For individual CDRs, we use Amino Acid Recovery (AAR,$\%$), representing the percentage of positions where the designed and native CDR sequences share the same amino acid, and Root Mean Square Deviation (RMSD), which measures the structural deviation of $C_\alpha$ atoms between the designed and native antibodies within the CDR region.

\textbf{Results}
We fix the antibody framework sequence and simultaneously design the sequences of all six CDR regions along with the complete antibody structure. We generate 10 antibody samples for each antigen.
Table \ref{tab2} shows the overall evaluation results for full antibody design. Both variants of AAMFM demonstrate strong performance compared to existing state-of-the-art models. Notably, AAMFM-CalDPO achieves the best scores across all four key metrics-Pll, AF3-score, pTM, and ipTM—highlighting its strength in generating sequences that are functionally sound, plausible, and antigen-specific. %

Table \ref{tab1} presents a fine-grained evaluation of each CDR region. AAMFM achieves state-of-the-art performance in 6 out of 12 subregions, with AAMFM-SFT and AAMFM-CalDPO ranking first in 4 and 2 subregions, respectively. This demonstrates AAMFM's strong capability in capturing local structural and sequence features across both heavy- and light-chain CDRs, producing designs that closely match the natural antibody distribution.
\vspace{-3mm}
\subsection{Epitope-binding CDR-H3 Co-design}
\begin{table}[t]
    \centering
    \caption{Results of CDR-H3 antibody design on SAbDab dataset}
    \begin{tabular}{cccccccc}
    \toprule
         Method&  Pll$\uparrow$ &AF3-score$\uparrow$ &A-RMSD$\downarrow$ &RMSD$\downarrow$&AAR$\uparrow$ & PHR$\downarrow$ & ipTM $\uparrow$\\
    \midrule
     ESM3-open& -1.88&  0.854& 2.02&3.32 & 33.04& \textbf{0.392}& 0.848\\
         Diffab& -1.69& 0.863 &2.12 & 2.99& 32.90& 0.443&0.856\\
 dyMEAN& -1.57&  0.863&1.69 &3.32 &39.49 &0.433 & 0.857\\
GeoAB & -1.60&  0.856&\textbf{1.41} &\textbf{1.67} & 40.40& \uline{0.413}&0.850\\
AbX  &-1.46 &0.877 &2.06 &3.06 &\textbf{43.14} &0.509 &0.871 \\
 
\midrule
 \textbf{AAMFM-SFT}& \uline{-1.20}&\uline{0.879} & \uline{1.55}&\uline{2.43} & \uline{40.55}&0.432 & \uline{0.878} \\
 
    \textbf{AAMFM-CalDPO} & \textbf{-1.18}&\textbf{0.894 }& 1.66 &2.62 &39.91 &0.438 &\textbf{0.890 }\\

   \bottomrule
 \end{tabular}

    \label{tab3}
\end{table}

\textbf{Setup, Baselines and Metrics} The overall experimental setup, baselines, and evaluation metrics largely same as those described in Section \ref{codesign}. This experiment differs in three aspects. First, only the CDR-H3 sequence is designed, while the full antibody structure is still generated. Second, we introduce an additional baseline, GeoAB \citep{lin2024geoab}, a graph model specifically designed for CDR-H3 co-design. Third, to focus more on local CDR-H3, we augment the evaluation metrics by modifying the calculation of the PLL score to apply only within the CDR-H3 region \citep{ye2024proteinbench}. 

\begin{wrapfigure}[12]{r}{0.6\textwidth}
     \centering
        \vspace{-4mm}
        \includegraphics[width=0.5\textwidth]{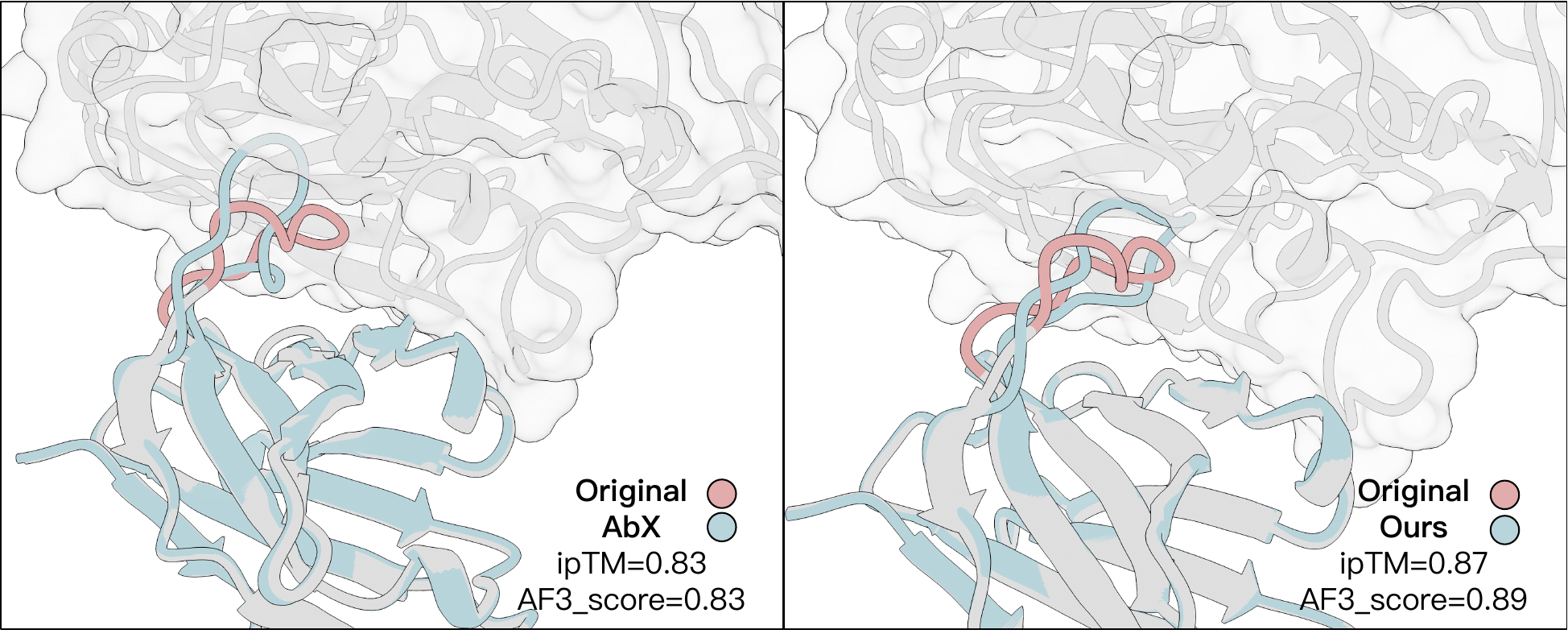}
        \vspace{-3mm}
        \caption{Visualization of the AF3 predicted structure of AbX and AAMFM-CalDPO designed CDR-H3 sequence. AAMFM achieves higher ipTM and AF3 score, highlighting its capability of functional design.}
        \label{case}
\end{wrapfigure}
\textbf{Results} In this setting, we employ two variants of our model to design only the CDR-H3 sequence  and the full antibody structure. Table \ref{tab3} presents a comparison between our models and baseline methods across several evaluation metrics. 
Our models consistently outperform state-of-the-art approaches in terms of sequence plausibility (Pll) and functional quality (AF3-score and ipTM). They also exhibit competitive performance on other metrics such as AAR and RMSD, further demonstrating their effectiveness in viable and realistic antibody design. Furthermore, we include a representative case study (Figure~\ref{case}) comparing the antibody CDR-H3 designed by AAMFM and AbX. This example highlights that AAMFM produces CDR-H3 sequences with superior functionality.

\subsection{Ablation and Analysis}
\label{sec:ablation}

\textbf{Setup and Metrics} The experimental setup and evaluation metrics follow those described in Section~\ref{codesign}. We identify four key components in our pipeline: SFT on antibody-antigen complex data, the antigen geometric-epitope-aware adapter, epitope conditioning, and Cal-DPO preference optimization. To assess the contribution of each component, we compare the full AAMFM-CalDPO model against the following ablation variants: (1) pretraining on OAS without subsequent SFT (\textbf{w/o SFT}); (2) performing SFT on SAbDab without the antigen adapter (\textbf{w/o adapter}); (3) disabling epitope input at inference by replacing the epitope mask with an all-zero tensor (\textbf{w/o epitope}); and (4) omitting the Cal-DPO stage and evaluating the SFT model directly (\textbf{w/o DPO}). For evaluation, we report PLL, AF3-score, and ipTM to assess overall sequence plausibility, foldability, and antigen-binding functionality, PHR to measure hydrophobicity, and CDR-H3-specific AAR and RMSD to evaluate local sequence recovery and structural accuracy.

\textbf{Results} As shown in Table~\ref{tab4}, we design all six CDR regions along with the full antibody structure to evaluate the contribution of each component. Experimental results show that each component plays a distinct role in the overall pipeline. 
Although removing SFT still yields competitive structural metrics, it results in less stable downstream DPO optimization, confirming its essential role in adapting the model to antibody-antigen interaction patterns before preference optimization.
Excluding the antigen adapter results in the most significant functional degradation, highlighting its critical role in geometric conditioning. 
Disabling epitope supervision degrades both foldability and antigen-binding confidence, while sequence recovery changes only marginally, suggesting that epitope conditioning primarily steers structural and functional compatibility rather than sequence identity.

Although SFT encourages the model to memorize the native antibody distribution—reflected in higher AAR and lower CDR-H3 RMSD—it inherently constrains exploration of the functional sequence space. Cal-DPO addresses this by shifting the optimization objective from sequence memorization toward preference-guided generation, enabling broader exploration of the functional sequence space. The modest decrease mirrors the distributional shift commonly observed in RL-based post-training, representing an expected trade-off rather than a regression.
Overall, Cal-DPO consistently improves sequence plausibility, foldability, and antigen-binding confidence, confirming that preference optimization effectively steers generation toward functional objectives.

\begin{table}[]
\caption{Ablation on CDRs design}
    \centering
        \begin{tabular}{ccccccc}
    \toprule
         Method & Pll $\uparrow$ & AF3-score $\uparrow$ & ipTM $\uparrow$ & RMSD $\downarrow$ & AAR $\uparrow$ & PHR $\downarrow$\\
    \midrule
ESM3-open    & -1.40          & 0.846          & 0.839          & 3.31           & 32.19          & \textbf{0.400} \\
w/o SFT      & \uline{-0.89}  & 0.876          & 0.869          & \textbf{2.49}  & 39.59  & 0.445 \\
w/o adapter  & -0.94          & 0.868          & 0.862          & 2.57           & 39.22          & 0.443 \\
w/o epitope  & -0.91             & \uline{0.883} & \uline{0.870}  & 2.66           & 38.98          & \uline{0.437} \\
\midrule
w/o DPO (SFT)    & -0.91      & 0.870          & 0.865          & \uline{2.54}   & \textbf{40.32} & 0.465 \\
AAMFM-CalDPO & \textbf{-0.87} & \textbf{0.892}  & \textbf{0.888} & 2.65           & \uline{39.88}          & 0.443 \\
    \bottomrule
    \end{tabular}
    \label{tab4}
\end{table}

\section{Conclusion and Limitations}
In this work, we present an antigen-specific antibody multi-modal foundation model AAMFM for functional design. This model leverages a powerful pre-trained protein language base model (ESM3) and uniquely integrates antibody sequence, structure, and specific antigen information, including geometric features and epitope sites, into a unified representation space using a lightweight adapter. Furthermore, AAMFM incorporates a novel finetune stage using Cal-DPO, leveraging preference signals extracted from a strong structural prior (AF3),  thereby aligning the generation process with key objectives of functional and plausible antibody design. Experimental results demonstrate that AAMFM achieves state-of-the-art performance across multiple antibody design tasks.

The main limitation of this work is that the designed antibodies have not yet been fully validated through in vitro experiments while we are actively collaborating with experimental research groups to pursue this. Another key future direction is to generalize the AAMFM framework beyond antibodies, extending it to the design of a broader class of protein complexes.

\newpage
\bibliographystyle{plain}

\normalem
\bibliography{references}

\newpage
\appendix

\section{Experiment Details}
\subsection{Baseline Details}

\textbf{DiffAb} \citep{luo2022antigen} We used DiffAb from the official GitHub repository (\href{https://github.com/luost26/diffab}{https://github.com/luost26/diffab}). To ensure a fair comparison, we retrained DiffAb on the same dataset as our model, using the configuration file \texttt{codesign\_multicdrs.yml} for 6 CDRs design and \texttt{codesign\_single.yml} for CDR-H3 design.

\textbf{dyMEAN} \citep{kong2023end} We used dyMEAN from the official GitHub repository (\href{https://github.com/THUNLP-MT/dyMEAN}{https://github.com/THUNLP-MT/dyMEAN}). For consistency, dyMEAN was retrained on our dataset using the configuration file \texttt{multi\_cdr\_design.json} for 6 CDRs design and \texttt{single\_cdr\_design.yml} for CDR-H3 design..

\textbf{AbX} \citep{zhou2024antigen} We used AbX from the official GitHub repository  (\href{https://github.com/CarbonMatrixLab/AbX}{https://github.com/CarbonMatrixLab/AbX}). The provided checkpoints(\href{https://zenodo.org/records/14577013}{https://zenodo.org/records/14577013}) were used for evaluation.

\textbf{GeoAB} \citep{lin2024geoab} We used GeoAB from the official GitHub repository (\href{https://github.com/Edapinenut/GeoAB}{https://github.com/Edapinenut/GeoAB}). To ensure a fair comparison, GeoAB was retrained on the same dataset as our model and only can perform CDR-H3 codesign.

\subsection{Metric Details} 
\textbf{Pll} Pll is calculated by AntiBERTy \citep{ruffolo2021deciphering} to gauge sequence plausibility. For CDR-H3, we modify the calculation of the Pll to apply only within the CDR-H3 region same as previous work\citep{ye2024proteinbench} for fine-grained comparison. AntiBERTy can be accessed at \href{https://github.com/jeffreyruffolo/AntiBERTy}{https://github.com/jeffreyruffolo/AntiBERTy}

\textbf{AF3-related scores} We adopted Protenix \citep{bytedance2025protenix} for evaluation due to its fully open-source implementation and its ability to extract evolutionary features using ESM2-3B embeddings, thereby avoiding the time-consuming MSA search. Specifically, we used version 0.4.0 from the \texttt{constraint\_esm} branch of the GitHub repository (\href{https://github.com/bytedance/Protenix/tree/constraint_esm}{https://github.com/bytedance/Protenix/tree/constraint\_esm}) to evaluate the sequences. The evaluation metrics include \textbf{pTM}, \textbf{ipTM}, and the \textbf{ranking score}.

\textbf{Proportion of Hydrophobic Residues (PHR)} Following previous works \citep{zhou2024antigen,ye2024proteinbench}, we use PHR to assess the antibodies specificity designed by the models. This metric is calculated only at CDR-H3 region. PHR is defined as the proportion of hydrophobic amino acids—specifically Alanine (A), Valine (V), Isoleucine (I), Leucine (L), Methionine (M), Phenylalanine (F), Tryptophan (W), and Tyrosine (Y). The lower PHR means better specificity.

\subsection{Dataset Preparation} For the synthetic dataset, we used a subset of the OAS database, which includes approximately 1.4 million paired antibody sequences predicted by IgFold (\href{https://data.graylab.jhu.edu/OAS_paired.tar.gz}{https://data.graylab.jhu.edu/OAS\_paired.tar.gz}, \href{https://data.graylab.jhu.edu/Jaffe2022.tar.gz}{https://data.graylab.jhu.edu/Jaffe2022.tar.gz}) and ABodyBuilder (\href{https://zenodo.org/records/7258553}{https://zenodo.org/records/7258553}).
We added Gaussian noise with a standard deviation of 0.1\,\AA{} to the atomic coordinates of all predicted protein structures.
For the DPO dataset, we used AAMFM-SFT to generate sequences based on all PDB structures in the RAbD dataset. We sampled sequences across different temperatures ranging from 0.1 to 1.5, generating 30 CDR-region sequences per complex at each temperature. Each generated sequence was then evaluated using both AntiBERTy and Protenix to obtain scoring metrics for downstream finetune.

\subsection{Implementation Details}
Our model was developed and implemented using the PyTorch framework. All experiments are run on a single A800 GPU, with a memory storage of 80GB. We use ESM-3 (version 3.1.6) as the base model, specifically the open-source 1.4B parameter version available at \href{https://github.com/evolutionaryscale/esm}{https://github.com/evolutionaryscale/esm}.

In the first training stage (on the OAS subset), we used the AdamW optimizer with a learning rate of 5e-5 and a batch size of 8. Mixed-precision training with bfloat16 (bf16) was employed to accelerate training and reduce memory consumption. We trained the model for 2 epochs.

In the second stage (Supervised Fine-Tuning, SFT), both the antigen and antibody sequences were fed into the model. During this phase, we trained additional adapter layers. To prevent overfitting, we froze the base model and trained only the adapters for the first 40 epochs using the AdamW optimizer with a learning rate of 1e-4 and a warm-up over 10\% of the training steps. After 40 epochs, we unfroze the base model and continued training for an additional 5 epochs with a learning rate of 1e-5 using AdamW (the adapter's learning rate is still 1e-4). Training was also conducted with bf16 mixed precision. Due to the increased input sequence length in this stage, we employed gradient accumulation with an effective batch size of 8 (batch size of 1 with accumulation steps set to 8).

In the third stage (Cal-DPO), we used the AdamW optimizer with a learning rate of 5e-5 and trained the model for 4 epochs, using bf16 mixed precision. Due to the increased input sequence length in this stage, we employed gradient accumulation with an effective batch size of 8 (batch size of 1 with accumulation steps set to 8).

\section{Additional Experimental Analysis}
\begin{figure}
    \centering
    \includegraphics[width=0.95\linewidth]{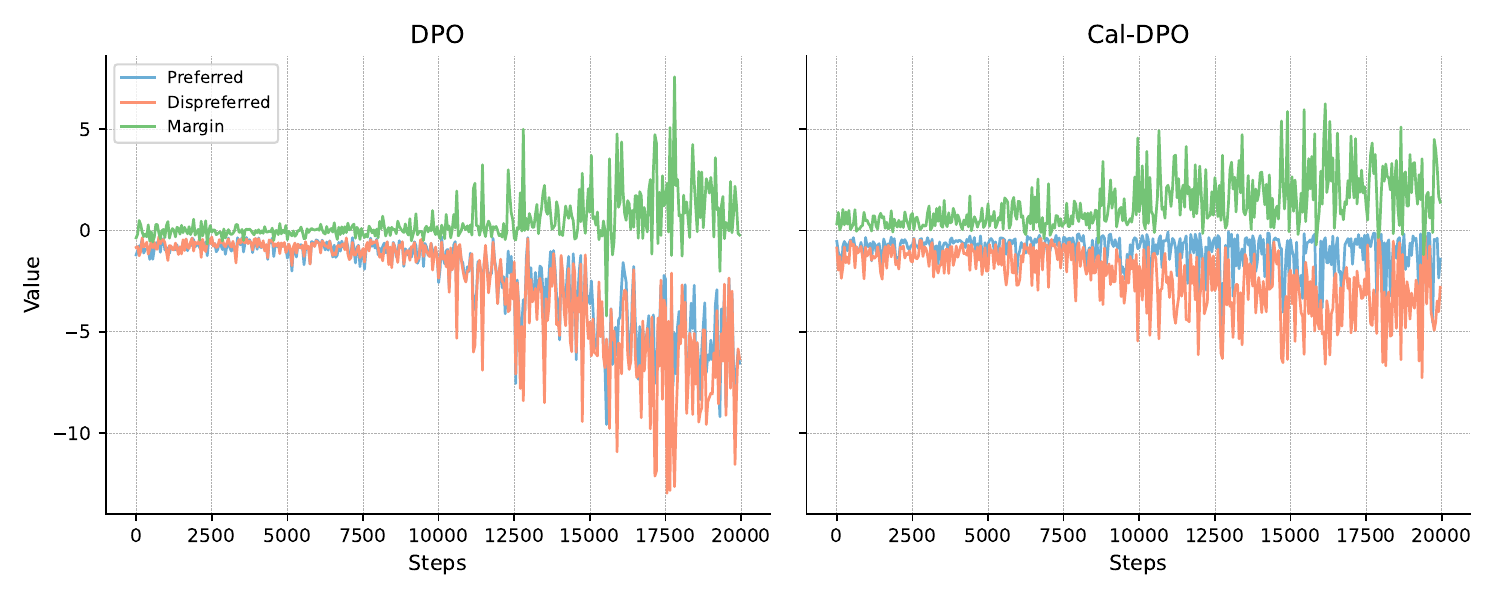}
    \caption{Training dynamic comparison between DPO and Cal-DPO}
    \label{c}
\end{figure}
We compared the training process of DPO and Cal-DPO algorithms, as illustrated in Figure~\ref{c}. While both methods exhibit an increasing margin (the difference between the log probabilities of preferred and dispreferred samples), we observed a key difference in their behavior. During standard DPO training, the log probabilities of both preferred and dispreferred sequences decrease simultaneously, indicating that the model may not fully capture the preference signal. In contrast, Cal-DPO maintains a relatively stable log probability for preferred sequences while decreasing the log probability of dispreferred ones. This behavior suggests that Cal-DPO more effectively enhances the model’s ability to favor preferred samples during generation.

\section{Detailed Related Work}
\textbf{Multi-modal Large Language Models}
Multimodal large language models (MLLMs) refer to LLMs trained on multiple modalities beyond text, enabling them to process or generate diverse data types such as images and videos. In the visual domain, models like LLaVA \citep{liu2023visual,liu2024llava} connect a vision encoder with an LLM to support general-purpose visual and language understanding. Similar strategies have been developed for multimodal video understanding \citep{song2024moviellm,tang2025video}. MLLMs have also been proposed for embodied AI and robotics \citep{xiong2024aic}, medical applications \citep{xu2024mlevlm}, and autonomous driving \citep{zhao2025sce2drivex}.

In the protein modeling domain, several recent works have introduced structural modalities to enhance sequence-only models, improving their representation and generative capabilities. Notable examples include SaProt \citep{su2023saprot}, LMDesign \citep{zheng2023structure}, ESM3 \citep{hayes2025simulating}, DPLM-2 \citep{wangdplm}, and S$^2$ALM \citep{yin2024s}. Inspired by this trend, AAMFM adopts a similar strategy to enhance multimodal antibody representations and improve the functionality of antibody design.

\textbf{Direct Preference Optimization}
Direct Preference Optimization (DPO) methods \citep{rafailov2023direct} simplifies RLHF \citep{ouyang2022training} by using contrastive learning on preference data, and has proven effective. However, the standard DPO algorithm has certain limitations: it focuses solely on relative preferences, disregarding the absolute values of both preferred and dispreferred samples, and heavily relies on the quality of preference-pair data \citep{xiao2024comprehensive}. To address these issues, a series of improved algorithms have been proposed \citep{amini2024direct,ethayarajh2024kto,meng2024simpo,liu2024tis}. IPO \citep{azar2024general} mitigates overfitting by introducing an additional loss term. IRPO \citep{pang2024iterative} explicitly incorporates the negative log-likelihood (NLL) of preferred samples into the loss, thereby increasing the probability of generating preferred outputs. Similarly, DPOP \citep{pal2024smaug} and AIPO \citep{shen2024aipo} add extra loss terms based on the Bradley-Terry model to prevent the probability of preferred samples from decreasing during training.

In contrast to these methods that only emphasize the preferred samples, Cal-DPO \citep{xiao2024cal} simultaneously considers the absolute values of both preferred and dispreferred samples. It enables the model to learn implicit rewards in contrastive preference learning, aligning model outputs with ground-truth rewards in accordance with human preferences. Owing to these advantages, we adopt Cal-DPO as an improved method to better align AAMFM with functional and plausible antibody design preferences.

\section{Method Details}
\subsection{Model Details}

We use ESM3-1.4B as our base model~\citep{hayes2025simulating}. ESM3 is a bidirectional transformer architecture that integrates multimodal information from protein sequence, structure, and function. These modalities are first embedded and fused at the input layer, then processed through a stack of transformer blocks. At the output, shallow multi-layer perceptron (MLP) heads project the final hidden states into token probabilities corresponding to each modality.
Instead of using modality-specific architectural components, ESM3 adopts a unified tokenization approach to represent the complexity of proteins in a shared multimodal feature space. This design enables efficient and scalable training. In particular, protein structures are tokenized using a discrete autoencoder that compresses 3D spatial information into discrete structural tokens.

To incorporate antigen geometry and antigen-antibody interface features for antigen-specific antibody generation, we introduce the Antigen Geometric-Epitope-aware Adapter.
Antigen 3D coordinates are first passed through a GNN (GearNet \citep{zhangprotein}) to obtain node embeddings, which are used as geometric features. These embeddings are projected into a unified feature space through a combination of MLPs, convolutional layer, and pooling layer. 
Meanwhile, epitope information is encoded as a binary matrix aligned with the antigen sequence (1 for epitope residues, 0 for other region), which is then passed through a learnable embedding layer that maps the binary values to an 8-dimensional space. These embeddings are combined with positional encodings and input into a Transformer encoder to obtain contextualized epitope representations.
The resulting geometric and epitope features are concatenated with the last-layer output of ESM3's base transformer (after projecting it from 1536 to 64 dimensions). The concatenated features are then mapped back to the original 1536-dimensional space and added to the base model's final hidden states.

\subsection{Direct Preference Optimization Details}
Given a preference pair $(y_w, y_l)$ under input context $c$ (e.g., masked antibody sequence $s_{fr}$, antigen features $G_{ag}$, and interface information $I$), the Bradley–Terry (BT) model assumes a latent reward function $r(y|c)$ and models the preference as:

\begin{equation}
P(y_w \succ y_l \mid c) = \frac{\exp(r(y_w \mid c))}{\exp(r(y_w \mid c)) + \exp(r(y_l \mid c))}.
\label{eq:bt}
\end{equation}

In the standard reinforcement learning from human feedback (RLHF) framework~\citep{ouyang2022training}, a reward model $r_\phi$ is trained by maximizing the expected reward under the optimized policy $\pi_\theta$, while regularizing it towards the reference policy $\pi_{\text{ref}}$ with a KL-divergence:

\begin{equation}
\max_{\pi_\theta} \ \mathbb{E}_{y \sim \pi_\theta} \left[ r_\phi(y \mid c) \right] - \beta D_{\text{KL}}(\pi_\theta(y \mid c) \| \pi_{\text{ref}}(y \mid c)).
\label{eq:rlhf}
\end{equation}

The optimal solution for the reward model has a closed form:

\begin{equation}
    \pi^*_\theta(y \mid c) = \frac{\pi_{\text{ref}}(y \mid c) \cdot \exp(r^*(y \mid c)/\beta)}{Z}
\end{equation}

where $Z$ is a normalization constant. This leads to the optimal reward model:

\begin{equation}
r^*(y \mid c) = \beta \log \frac{\pi_\theta(y \mid c)}{\pi_{\text{ref}}(y \mid c)} + \beta \log Z.
\label{eq:optimal_reward}
\end{equation}

Substituting Equation~\ref{eq:optimal_reward} into the Bradley-Terry model in Equation~\ref{eq:bt} and optimizing the log-likelihood, the Direct Preference Optimization (DPO) objective~\citep{rafailov2023direct} becomes:

\begin{equation}
\mathcal{L}_{\text{DPO}}(\theta; \pi_{\text{ref}}) = - \mathbb{E}_{(c, y_w, y_l) \sim \mathcal{D}} \left[
\log \sigma\left( \beta \log \frac{\pi_\theta(y_w \mid c)}{\pi_{\text{ref}}(y_w \mid c)} \right)
- \log \sigma\left(  \beta \log \frac{\pi_\theta(y_l \mid c)}{\pi_{\text{ref}}(y_l \mid c)} \right)
\right],
\label{eq:dpo_final}
\end{equation}

where $\sigma(\cdot)$ is the sigmoid function and $\beta$ is a temperature parameter controlling preference sharpness.

Cal-DPO introduces an additional calibration loss term. Defining the target margin $M = 1 / (2\beta)$, the calibration loss encourages the log-likelihood ratios to align with this margin:
\begin{equation}
\mathcal{L}_{\text{Cal}}(\theta; \pi_{\text{ref}}) = \mathbb{E}_{(c, y_w, y_l) \sim \mathcal{D}} \left[ ( \log \sigma( \beta \log \frac{\pi_\theta(y_w \mid c)}{\pi_{\text{ref}}(y_w \mid c)}  - M \right)^2 + \left( \log \sigma\left( \beta \log \frac{\pi_\theta(y_l \mid c)}{\pi_{\text{ref}}(y_l \mid c)} + M \right)^2 \right]
\end{equation}
\vspace{-3mm}

The final training loss is a weighted sum of the DPO and calibration terms:
\vspace{-3mm}

\begin{equation}
\mathcal{L}_{\text{Cal-DPO}} = \mathcal{L}_{\text{DPO}} + \lambda \cdot \mathcal{L}_{\text{Cal}}
\end{equation}

\section{Broader Impacts} Our work has broad potential applications in biotechnology and therapeutic development. For instance, the ability to design functional antibodies that bind to specific antigens could contribute to the development of vaccines, or the engineering of agonists and antagonists that modulate specific immune pathways or signaling proteins. Such applications have the potential to advance precision medicine, immunotherapy, and the treatment of various diseases.

We explicitly prohibit the use of our model for potentially harmful purposes, such as the development of biological weapons, targeting of human proteins for malicious intent, or any application that may pose significant ethical, safety, or biosecurity risks.

\section{Codes}
 Our code is available at \url{https://github.com/XL-S224/AAMFM}.

\end{document}